\documentclass{IEEEtran4PSCC}

\usepackage[caption=false,font=footnotesize]{subfig}	
\usepackage[dvips]{graphicx}

\usepackage{algorithm}
\usepackage{algorithmicx}
\usepackage{algpseudocode}
\usepackage{courier}

\usepackage{amsmath,amsfonts,amssymb,amsthm}
\usepackage{mathtools}
\usepackage{mathrsfs}	
\usepackage{subdepth}	

\usepackage{array}	
\usepackage[inline]{enumitem} 

\usepackage{hyperref}
\usepackage[sort,compress,capitalise]{cleveref}	
\usepackage{url}

\usepackage{xcolor} 
\usepackage{xspace} 
\usepackage{lipsum}

\usepackage{tikz}
\usetikzlibrary{babel}  
\usepackage[straightvoltages,nooldvoltagedirection]{circuitikz}
\usetikzlibrary{shapes,arrows}
\usepackage{verbatim}
\usepackage{multirow}

 \usepackage{booktabs}

\DeclareMathOperator*{\argmin}{arg\,min}
\DeclareMathOperator*{\argmax}{arg\,max} 

{
\theoremstyle{plain}

}

{
\theoremstyle{definition}

}

\crefname{Hypothesis}{Hyp.}{Hyps.}
\Crefname{Hypothesis}{Hyp.}{Hyps.}

\crefname{Lemma}{Lemma}{Lemmata}
\Crefname{Lemma}{Lemma}{Lemmata}

\crefname{Definition}{Def.}{Defs.}
\Crefname{Definition}{Def.}{Defs.}

\newcommand{\DFT}[1][]{DFT\xspace}  

\usepackage{bm}
\usepackage{mathtools}

\DeclarePairedDelimiter\floor{\lfloor}{\rfloor} 
\makeatletter
\let\old@ps@headings\ps@headings
\let\old@ps@IEEEtitlepagestyle\ps@IEEEtitlepagestyle
\def\psccfooter#1{%
    \def\ps@headings{%
        \old@ps@headings%
        \def\@oddfoot{\strut\hfill#1\hfill\strut}%
        \def\@evenfoot{\strut\hfill#1\hfill\strut}%
    }%
    \def\ps@IEEEtitlepagestyle{%
        \old@ps@IEEEtitlepagestyle%
        \def\@oddfoot{\strut\hfill#1\hfill\strut}%
        \def\@evenfoot{\strut\hfill#1\hfill\strut}%
    }%
    \ps@headings%
}
\makeatother

\psccfooter{%
        \parbox{\textwidth}{\hrulefill \\ \small{22nd Power Systems Computation Conference} \hfill \begin{minipage}{0.2\textwidth}\centering \vspace*{4pt} \includegraphics[scale=0.06]{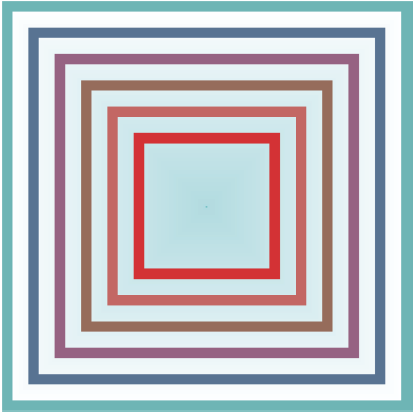}\\\small{PSCC 2022} \end{minipage} \hfill \small{Porto, Portugal --- June 27 -- July 1, 2022}}%
}

\begin{document}
\title{Optimal Grid-Forming Control of Battery Energy Storage Systems Providing Multiple Services: Modelling and Experimental Validation}

\author{
\IEEEauthorblockN{Francesco Gerini, Yihui Zuo, Rahul Gupta, \\Rachid Cherkaoui, Mario Paolone}
\IEEEauthorblockA{Distributed Electrical Systems Laboratory (DESL)\\
Ecole Polytechnique Fédérale de Lausanne, EPFL\\
Lausanne, Switzerland\\
\{francesco.gerini, yihui.zuo, rahul.gupta, \\rachid.cherkaoui, mario.paolone\}@epfl.ch}
\and
\IEEEauthorblockN{Elena Vagnoni}
\IEEEauthorblockA{Technology Platform for Hydraulic Machines (PTMH)\\
Ecole Polytechnique Fédérale de Lausanne, EPFL\\
Lausanne, Switzerland\\
\{elena.vagnoni\}@epfl.ch}
}

\maketitle
\begin{abstract}

This paper proposes and experimentally validates a joint control and scheduling framework for a grid-forming converter-interfaced BESS providing multiple services to the electrical grid.
The framework is designed to dispatch the operation of a distribution feeder hosting heterogeneous prosumers according to a dispatch plan and provide frequency containment reserve and voltage control as additional services. 
The framework consists of three phases. In the day-ahead scheduling phase, a robust optimization problem is solved to compute the optimal dispatch plan and frequency droop coefficient, accounting for the uncertainty of the aggregated prosumption. In the intra-day phase, a model predictive control algorithm is used to compute the power set-point for the BESS to achieve the tracking of the dispatch plan. Finally, in a real-time stage, the power set-point originated by the dispatch tracking is converted into a feasible frequency set-point for the grid forming converter by means of a convex optimisation problem accounting for the capability curve of the power converter. The proposed framework is experimentally validated by using a grid-scale 720 kVA/560 kWh BESS connected to a 20 kV distribution feeder of the EPFL hosting stochastic prosumption and PV generation.

\end{abstract}

\thanksto{\noindent Submitted to the 22nd Power Systems Computation Conference (PSCC 2022).}
\vspace{-0.4cm}


\section{Introduction}
\label{Section:Intro}
\subsection{Motivation}
Power systems are evolving towards environmentally sustainable networks by increasing the share of renewable power generation interfaced by power converters. 
The decline of grid inertia levels and a lack of frequency containment delivered by conventional synchronous generators~\cite{CAISOfreq1,ERCOTinertia}, in addition to the uncertainty associated to the forecast of renewable resources, require restoring an adequate capability of regulating power to assure reliable operation of interconnected power systems~\cite{AEMOrenint1}. 

In this context, network operators are motivated to set hard requirement on the dispatchability of connected resources  and to incorporate assets with high ramping capability to maintain frequency containment performance~\cite{ERCOT_rehman2018dynamic,AEMO_operatormaintaining}. 
An emerging concept to tackle the challenge of dispatchability of power distribution systems hosting stochastic power generation is to exploit the utility-scale battery energy storage systems (BESSs). Moreover, in recent years, BESSs are  increasingly deployed for frequency regulation, thanks to their large ramping capability, high round-trip efficiency, and commercial availability~\cite{SciBattery4Grid,BESSdeploy_cole2019cost}.
In this respect, control frameworks of BESS providing multiple ancillary services to the power system are of high interest  to fully take advantage of BESSs investments~\cite{Hornsdale2018initial,MEGEL201548}. 



The state-of-the-art  has presented optimal solution for ancillary services provision~\cite{namor_control_2019,sossan_achieving_2016,zecchino_optimal_2021}.
In particular, \cite{namor_control_2019} proposes to solve an optimization problem that allocates the battery power and energy budgets to different services in order to maximize battery exploitation. Nevertheless, to the authors' best knowledge, the dispatch tracking problem is oversimplified and does not ensure the BESS operation to be within the physical limits.
On the other hand, \cite{sossan_achieving_2016} tackles the problem of dispatching the operation of a cluster of stochastic prosumers through a two-stage process, which consists of a day-ahead dispatch plan determined by the data-driven forecasting and a real-time operation tracking the dispatch plan via adjusting the real power injections of the BESS with a \emph{Model Predictive Control} (MPC).
Finally, \cite{zecchino_optimal_2021} proposes a control method for BESSs to provide concurrent \emph{Frequency Containment Reserve} (FCR) and local voltage regulation services.

Despite the efforts, all the proposed solutions rely on \textit{grid-following} (GFL) control strategies, therefore ignoring the possibility of controlling the BESS converter in \textit{grid-forming} (GFR) mode. 
Even if the majority of converter-interfaced resources is currently controlled as grid-following units~\cite{PV_MPPT_GFL, CurrentgrieCSC2, CurrentgrieCSC1}, future low-inertia grids are advocated to host a substantiate amount of grid-forming units providing support to both frequency/voltage regulation and system stability~\cite{GFM_roadmap,reviewGFM2,GFMpotential_1}.
Recent studies have proved GFR control strategies to outperform GFL in terms of frequency regulation performance in low-inertia power grids~\cite{zuo_performance_2021}. Furthermore, the impact of GFR converters on the dynamics of a reduced-inertia grids has been investigated in~\cite{zuo_effect_2020}, which quantitatively proved the good performance of GFR units in limiting the frequency deviation and in damping the frequency oscillations in case of large power system contingencies. 
Nevertheless, the existing scientific literature lacks of studies assessing the performance of GFR units in supporting the frequency containment process of large interconnected power grids. Moreover, to the best of the authors knowledge, GFR units have never been proved able to provide services such as feeder dispatchability. In fact, studies on the GFR units synchronizing with AC grids are mostly limited to ancillary services provision and their validations is based on either simulation~\cite{zuo_performance_2021}-\cite{liu_comparison_2016} or to experiments on ideal slack buses with emulated voltage~\cite{qoria_pll-free_2020,rosso_robust_2019}.

\subsection{Paper's Contribution}
With respect to the existing literature discussed in the previous section, the contributions of this paper are the following.
\begin{itemize}
    \item The development of a control framework for GFR converter-interfaced BESS, tackling the optimal provision of multiple services, which relies on  existing grid-following control strategies \cite{namor_control_2019,sossan_achieving_2016,zecchino_optimal_2021}. The control framework for the simultaneous provision of feeder dispatchability, FCR and voltage regulation aims to maximize the battery exploitation in the presence of uncertainties due to stochastic demand, distributed generation, and grid frequency. 
\item The experimental validation of the proposed framework by using a 560 kWh BESS interfaced with a 720 kVA GFR-controlled converter to dispatch the operation of a 20 kV distribution feeder hosting both conventional consumption and distributed \emph{Photo-Voltaic} (PV) generation.
\item The performance assessment of the GFR-controlled BESS providing simultaneously dispatching tracking and FCR provision. In particular, the frequency regulation performance of the GFR-controlled BESS is evaluated and compared with the case of GFR only providing FCR and with the GFL case. 
\end{itemize}

The paper is organised as follows. \cref{Section:ProblemFormulation} proposes the general formulation of the control problem for the BESS providing multiple services simultaneously. \cref{Subsection:ControlFramework} presents a detailed description of the three-stage control framework.  \Cref{Section:Results} provides the validation of the proposed framework by means of real-scale experiments. Finally, \cref{Section:Conclusions} summarizes the original contributions and main outcomes of the paper and proposes perspectives for further research activities.
        
\section{Problem Statement}
\label{Section:ProblemFormulation}
In this paper, the dispatchability of distribution feeders and the simultaneous provision of FCR and voltage regulation is tackled by controlling a grid-forming converter-interfaced BESSs. Specifically, it is ensured the control of the operation of a group of prosumers (characterized by both conventional demand and PV generation that are assumed to be uncontrollable) according to a scheduled power trajectory at 5 minutes resolution, called dispatch plan, determined the day before operation. The day-ahead scheduler relies on a forecast of the local prosumption. 
The multiple-service-oriented framework consists of three stages, each characterised by different time horizons: 
\begin{enumerate*}[label=(\roman*)]
\item The dispatch plan is computed on the day-ahead (i.,e., in agreement with most common practice), where the feeder operator determines a dispatch plan based on the forecast of the
prosumption while accounting also for the regulation capacity of BESSs~\cite{dispatchparctice}. Regarding  the GFR droop computation, by referring to state-of-the-art practice of \emph{Transmission System Operators} (TSO) of France, Germany, Belgium, the Netherlands, Austria and Switzerland \cite{gestionnaire_du_reseau_de_transport_delectricite_rte_frequency_2020},  \cite{noauthor_regelleistung_nodate},  the FCR is supposed to be allocated on a daily basis. For these reasons, in the day-ahead stage, an optimization problem is solved to allocate the battery power and energy budgets to the different services by determining a dispatch plan at a 5-minute resolution based on the forecast of the prosumptiom and computing the droop for the FCR provision.
\item In the intermediate level stage, with a 5-minute horizon, the active power injections of the BESS are adjusted by means of a MPC targeting both the correction of the mismatch between prosumption and dispatch plan (as proposed by \cite{sossan_achieving_2016}) and the FCR provision. The MPC is actuated every 10 seconds to both ensure a correct tracking over the 5 minutes window and avoid overlapping the dispatch tracking with the FCR action.
\item In the final stage, computed each second, the MPC active power command is converted into a feasible frequency set-point for the GFR converter. 
As a matter of fact, the feasible PQ region of the BESS power converter is a function of the battery DC-link and AC-grid status~\cite{zecchino_optimal_2021}. For this reason, the feasibility of the grid-forming frequency reference set-point is ensured by solving every second an efficient optimization problem that takes into account the dynamic capability curve of the DC-AC converter and adjusts the set point accordingly. 
Eventually, the feasible frequency set-point is implemented in the GFR controller which intrinsically superposes the frequency control action on the active power dispatch. 
\end{enumerate*}

\section{Control Framework}
\label{Subsection:ControlFramework}

\subsection{Day-Ahead Stage}
\begin{figure}
    \centering
    \includegraphics[width=0.75\linewidth]{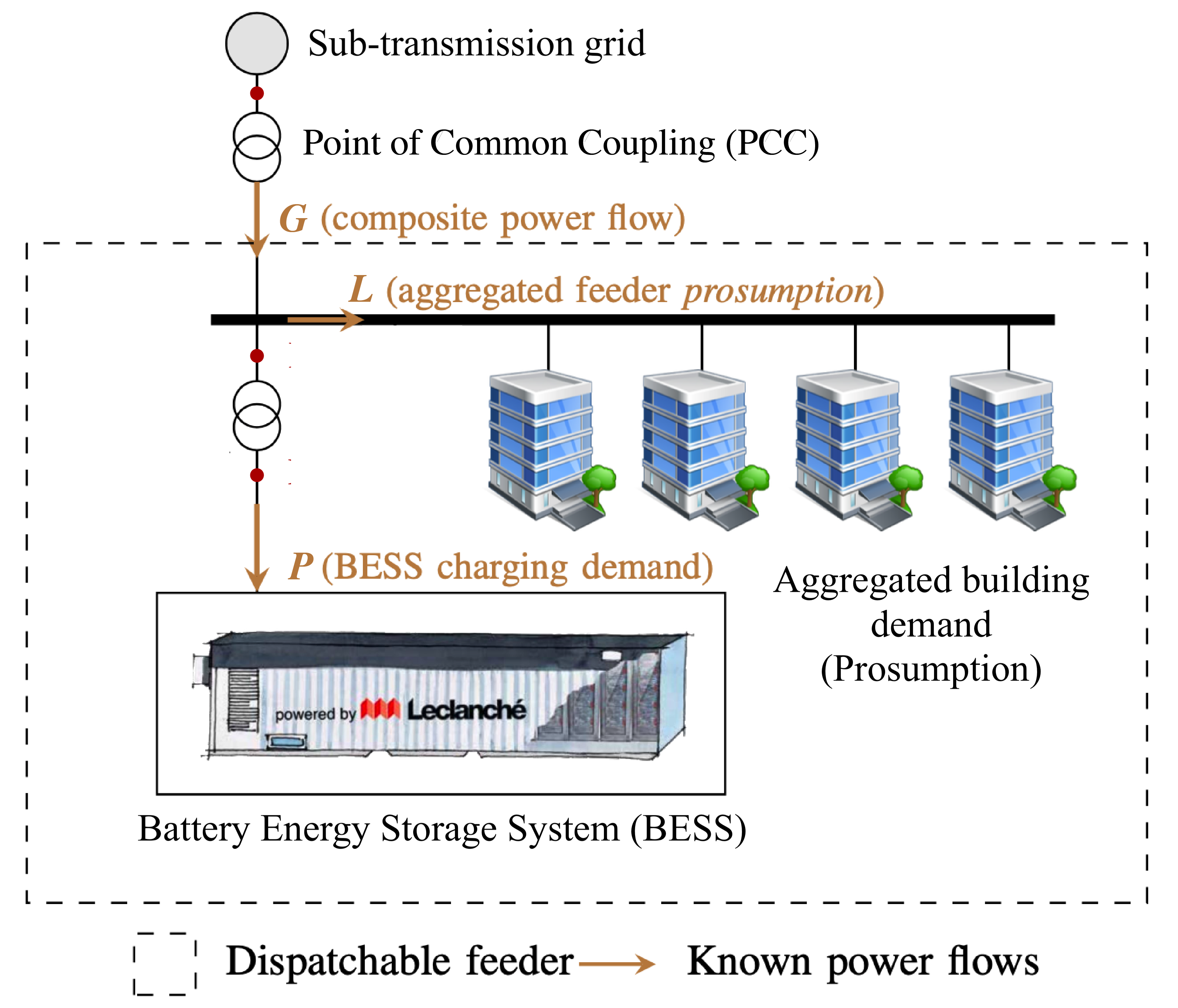}
    \caption{Feeder for problem statement}
    \label{fig:ExperimentalSetup}
\end{figure}

The objective of the day ahead is to compute a dispatch plan $\boldsymbol{\hat{G}}$ for a distribution feeder and to simultaneously contract with the TSO a certain frequency droop, for the BESS FCR provision. A representation of the feeder and the corresponding power flows is shown in Fig.~\ref{fig:ExperimentalSetup}.
The BESS bidirectional real power flow is denoted by $P$, while $G$ is the composite power flow as seen at the \emph{Point of Common Coupling} (PCC). The aggregated building demand is denoted by $L$ and, by neglecting grid losses, it is estimated as: 
\begin{equation}
    L=G-P
    \label{eq:PCCpower}
\end{equation}
Further to the achievement of feeder dispatchability, the proposed framework allows the GFR-controlled BESS to react against the grid-frequency variation in real time, i.e., providing FCR. The latter action is automatically performed by the converter operated in GFR mode, where the power flowing from to the BESS can be computed as:
\begin{align}
        P &= \sigma_f \cdot (f-f_{ref})
    \label{eq:GFRpower}\\
        Q &= \sigma_v \cdot (v-v_{ref})
    \label{eq:GFRQpower}
\end{align}

In \cref{eq:GFRpower} and \cref{eq:GFRQpower} $\sigma_f$ and $\sigma_v$\footnote{While the $\sigma_f$ is computed in the day-ahead problem for the FCR service, the $\sigma_v$ is considered to allow for adjusting reactive power in the real-time stage (see in~\cref{section:RTcontroller}) and is  determined according to the voltage control practice recommend in~\cite{noauthor_commission_nodate-2}.} are respectively the frequency and voltage droop fixed at the day-ahead stage, while $f_{ref}$ and $v_{ref}$ are the frequency and voltage reference set-point (i.e., real-time command) of the GFR converter. The target of the control problem is to regulate the composite power flow $G$ to respect the dispatch plan fixed on the day-ahead planning and the frequency containment action accorded with the TSO.

The formulation of the day-ahead problem considering the provision of both FCR and dispatchability by \cite{namor_control_2019} can be adapted to the grid-forming case. The mathematical formulation is hereby proposed:
\begin{subequations}\label{eq:dayahead}
\begin{alignat}{2}
 &\phantom{text{subject to:}}[\sigma_f^0, \boldsymbol{F}^o] = \!\argmin_{\sigma_f \in \mathbb{R}^+, \boldsymbol{F} \in \mathbb{R}^N} (\sigma_f)&             \label{eq:optProb}\\
 &\text{subject to:}& \nonumber\\
& {SOE_0}+\frac{1}{E_{\text{nom}}}\bigg[\frac{T}{N}\sum_{i=0}^n{(F_i+L_i^\uparrow)}+\sigma_f W_{f,n}^\uparrow \bigg]\le SOE_{\text{max}}, \label{eq:constraint1}  &  \\
& {SOE_0}+\frac{1}{E_{\text{nom}}}\bigg[ \frac{T}{N}\sum_{i=0}^n{(F_i+L_i^\downarrow)}+\sigma_f W_{f,n}^\downarrow \bigg] \ge SOE_{\text{min}}, \label{eq:constraint2}  & \\
& F_n + L_n^\uparrow + 0.2\sigma_f \ge P_{\text{max}}, \label{eq:constraint3}  & \\
& F_n + L_n^\downarrow + 0.2\sigma_f \le P_{\text{max}}, \label{eq:constraint4}  & 
\end{alignat}
\end{subequations}
where:
\begin{itemize}
    \item $T$ is the total scheduling time window (i.e., $T$ = 86400 seconds) discretized in $N$ time steps ($N=288$, i.e., the dispatch plan is divided into 5 minutes windows) and each step is denoted by the subscript $n$ with $n=0,...,N-1$. 
    \item $ \bm{\hat{L}} = \hat{L}_1,...,\hat{L}_{N}$ is  the forecast profile of feeder prosumption and {$ \bm{{F^o}} = F_1,... F_{N}$} is the BESS power offset profile which is computed to keep the BESS stored energy at a value capble to compensate for the difference between prosumers' forecasted and realized power. The day-ahead dispatch plan $\bm{\hat{G}}=\hat{G}_{1},...,\hat{G}_{N}$ is the sum of the above two terms, as in \cref{eq:PCCpower}.
    \item $\sigma_f$ is the FCR droop expressed in kW/Hz
    \item $W_{f,k}$ denotes the integral of frequency deviations over a period of time, and it represents the energy content of the signal given by the frequency deviation from its nominal value.
    \item The BESS limits in terms of \emph{State Of Energy} (SOE) and power are expressed respectively with $SOE_{\text{min}}$, $SOE_{\text{max}}$, $P_{\text{min}}$ and $P_{\text{max}}$, while $E_{\text{nom}}$ is the nominal BESS energy.
\end{itemize}
It is worth mentioning that the optimization problem described by \cref{eq:optProb,eq:constraint1,eq:constraint2,eq:constraint3,eq:constraint4} prioritizes the dispatchability of the feeder over the FCR provision on the day-ahead planning. This choice is nevertheless user-dependent, based on the economical convenience of the provided service or on grid core requirements. For example, if the user stipulates a contract with the TSO for FCR provision, this service can be prioritized, and the remaining energy can be allocated for the dispatch service, that will inevitably not be always achieved if the prosumption stochasticity is too high\footnote{Further discussion can be find in \cref{Section:Results}}.

\subsection{Intra-day Stage}
In the intra-day stage a MPC algorithm is used to target the fulfillment of the mismatch between average prosumption for each 5-minute period and dispatch plan plus FCR action accorded with the TSO for the same time-window.
Since the MPC action has a time-sampling of 10 seconds, the index $k = 0,1,2,...,K-1$ is introduced to denote the rolling 10 seconds time interval, where $K=8640$ is the number of 10 second periods in 24 hours.
The value of the prosumption set-point retrieved from the dispatch plan for the current 5-minute slot is indicated by the \textit{k}-index as: 
\begin{equation}
    G_k^* = \hat{G}_{\floor{\frac{k}{30}}}
\end{equation}
where $\floor{\cdot}$ denotes the nearest lower integer of the argument, and 30 is the number of 10-second interval in a 5-minute slot. The first and the last 10-second interval for the current 5-minutes are denoted as $\underline{k}$ and $\overline{k}$, respectively:
\begin{equation}
    \underline{k} = \floor{\frac{k}{30}}\cdot 30
\end{equation}
\begin{equation}
    \overline{k} = \underline{k} +30-1
\end{equation}

\begin{figure}[h!]
    \centering
    \includegraphics[width=0.9\linewidth]{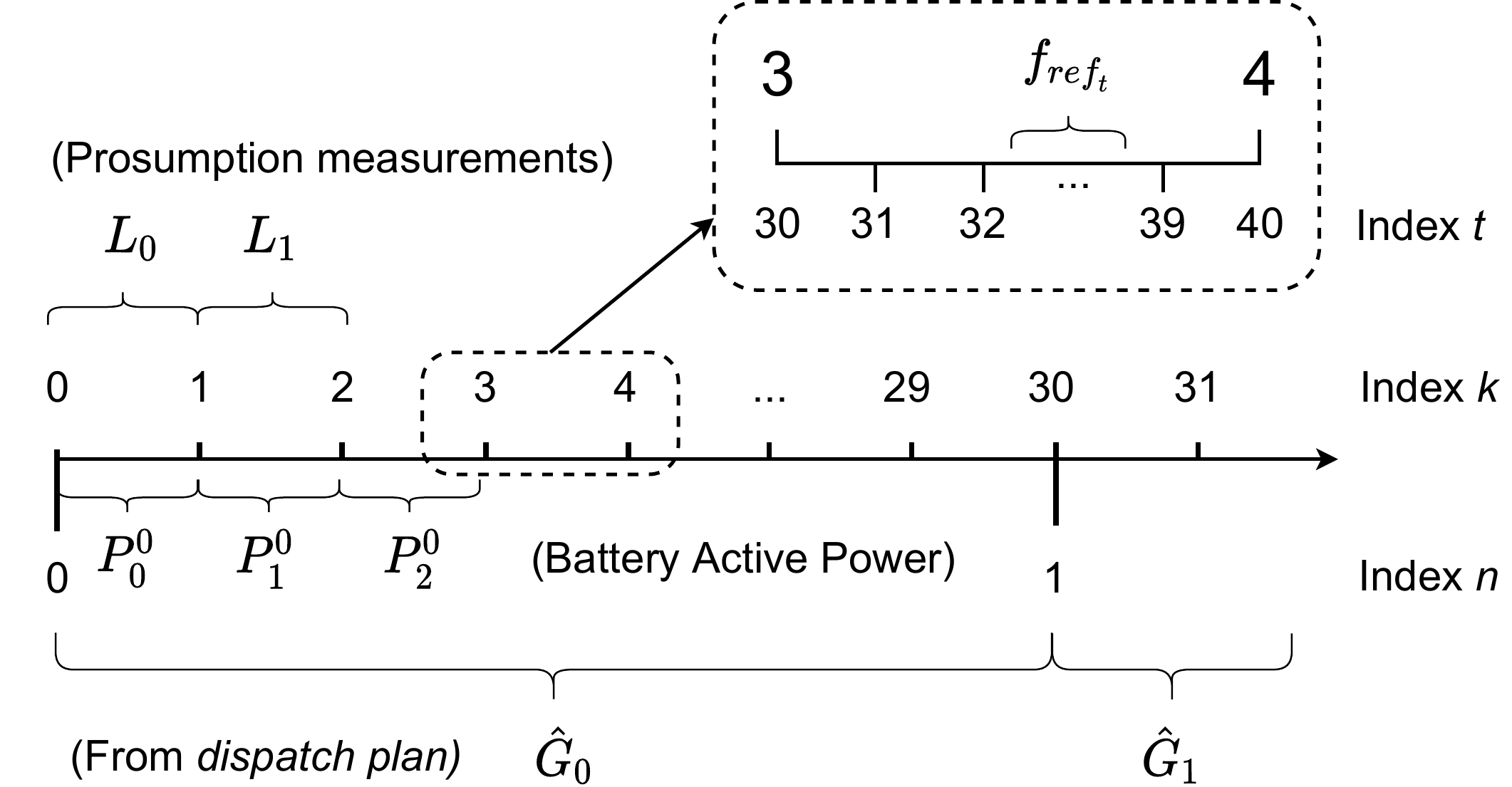}
    \caption{Timeline for the MPC problem}
    \label{fig:timeline}
\end{figure}

A graphical representation of the execution timeline for the MPC problem is given by Fig. \ref{fig:timeline} displaying the first thirty-one 10-seconds intervals of the day of operation. The figure shows the BESS power set-point $P_2^o$, which has been computed by knowing the prosumption realizations $L_0$ and $L_1$, and the average prosumption set-point to be achieved in the 5-minute interval (i.e., first value of the dispatch plan $\hat{G}_{0}$).
A similar control problem, not including the simultaneous provision of FCR by the BESS, is described in \cite{sossan_achieving_2016}. For this reason, the MPC problem proposed in \cite{sossan_achieving_2016} is modified as follows, to account for the provision of multiple services by means of GFR converter. 
Considering \cref{eq:PCCpower}, the average composite power flow at the PCC (prosumption + BESS injection) is given by averaging the available information until $k$ as:
\begin{equation}
G_k=\frac{1}{k-\underline{k}}\cdot \sum_{j=\underline{k}}^{k-1}(L_j+P_j)
    \label{eq:G_k}
\end{equation}
Then, it is possible to compute the expected average composite flow at the PCC at the end of the 5 minutes window as:
\begin{equation}
    G_k^+ = \frac{1}{30} \left( (k-\underline{k})\cdot G_k + \sum_{j=k}^{\overline{k}}{\hat{L}_{j|k}} \right) \label{eq:MPCobj}
\end{equation}
where a persistent\footnote{As shown in \cite{scolari2018comprehensive}, the persistent predictor performs well given the short MPC horizon time and fast control actuation in this application.} forecast is used to model future realizations, namely {$\hat{L}_{j|k} = L_{k-1}, j = k,...,\overline{k}$}.
The energy error between the realization and the target (i.e. dispatch plan plus FCR energy) in the 5-minute slot is expressed (in kWh) as: 
\begin{equation}
    e_k = \frac{300}{3600}\cdot (G_k^* - G_k^+ + \Delta G_k^F)
\end{equation}
where 300 s and 3600 s are the number of seconds in a 5 minutes interval and 1 hour interval, respectively. The additional term $\Delta G_k^F$ considers the deviation caused by the frequency containment response of the GFR converter:
\begin{equation}
    \Delta G_k^F = \frac{1}{30}\sum_{j=\underline{k}}^{k-1}(50-f_j) \cdot \sigma_f
\end{equation}
where $f_j$ is the frequency measurement at time $j$. 
Finally, the MPC can be formulated to minimize the error $e_k$ over the 5-minutes window, subject to a set of physical constraints such as BESS SOC, DC voltage and current operational limit. It should be noted that including the term $\Delta G_k^F$ in the energy error function fed to the MPC allows for decoupling the dispatch plan tracking with the frequency containment response provided by the BESS in each 5-minute slot. In particular, the omission of this term while operating the BESS in GFR mode, can create conflicts between dispatch plan tracking and frequency containment provision.
As in \cite{sossan_achieving_2016}, in order to achieve a convex formulation of the optimization problem, the proposed MPC problem targets the maximisation of the sum of the equally weighted BESS DC-side current values over the shrinking horizon from $k$ to $\overline{k}$ in its objective \eqref{eq:mpcProb} while constraining the total energy throughput to be smaller or equal to the target energy $e_k$.
The optimization problem is formulated as:
\begin{subequations}
\begin{alignat}{2}
& \bm{i}_{\overline{k}|k}^o = \!\argmax_{i\in\mathbb{R}^{(k-\overline{k}+1)}} (\bm{1}^T\bm{i}_{\overline{k}|k})       &      \label{eq:mpcProb}\\
\text{subject to:}\nonumber  
\\
& \alpha \bm{v}_{\overline{k}|k}^T \bm{i}_{\overline{k}|k} \leq e_k \label{eq:mpc1}& \\
&\bm{1}\cdot i_{min}\preceq \bm{i}_{\overline{k}|k}\preceq\bm{1}\cdot i_{max} & \label{eq:mpc2}\\
&\bm{1}\cdot \Delta i_{min}\preceq H\bm{i}_{\overline{k}|k}\preceq\bm{1}\cdot \Delta i_{max} & \label{eq:mpc3}\\
&\bm{v}_{\overline{k}|k} = \phi_v x_k +\psi_i^v\bm{i}_{\overline{k}|k}+\psi_1^v\bm{1}&\label{eq:mpc4}\\
&\bm{1}\cdot v_{min}\preceq\bm{v}_{\overline{k}|k}\preceq\bm{1}\cdot v_{max}&\label{eq:mpc5}\\
&\textbf{SOC}_{\overline{k}|k} = \phi^{\text{SOC}}\text{SOC}_k +\psi_i^{\text{SOC}}i_{\overline{k}|k}&\label{eq:mpc6}\\
&\bm{1}\cdot \text{SOC}_{min}\preceq\textbf{SOC}_{\overline{k}|k}\preceq\bm{1}\cdot \text{SOC}_{max} & \label{eq:mpc7}
\end{alignat}
\end{subequations}
where 
\begin{itemize}
    \item $\bm{i}_{\overline{k}|k}^o$ is the computed control action trajectory, $\bm{1}$ denotes the all-ones column vector, the symbol $\preceq$ is the component-wise inequality, and the bold notation denotes the sequences obtained by stacking in column vectors the realizations in time of the referenced variables, e.g. $\bm{v}_{\overline{k}|k} = [v_k,...v_{\overline{k}}]^T$. 
    \item In \eqref{eq:mpc1}, the BESS energy throughput (in kWh) on the AC bus is modeled as $E_{\overline{k}|k}(\cdot) = \alpha \bm{v}_{\overline{k}|k}^T \bm{i}_{\overline{k}|k}$, where $\bm{v}_{\overline{k}|k}$ and $\bm{i}_{\overline{k}|k}$ are the battery DC voltage and current, respectively, and $\alpha = 10/3600$ is a converting factor from average power over 10 seconds to energy expressed in kWh.  
    \item The inequality \eqref{eq:mpc2} and \eqref{eq:mpc3} are the constraints on the magnitude and rate of change for the BESS current, respectively. The matrix $H\in \mathbb{R}^{(k-\overline{k}+1)\times(k-\overline{k}+1)}$ is 
    \begin{equation}
        H = 
        \begin{bmatrix}
             1&-1 &0 &0 &...& 0 \\
             0&1 &-1 &0 &...& 0 \\
             \vdots&\vdots&\vdots&\vdots&\vdots&\vdots\\
             0&0 &0 &... &1& -1 
        \end{bmatrix}
    \end{equation}
    \item  The equality \eqref{eq:mpc4} is the \emph{Three-Time-Constant (TTC)} electrical equivalent circuit model of the voltage on DC bus, whose dynamic evolution can be expressed as a linear function of battery current by applying the transition matrices $\phi^v$,$\psi^v$,$\psi_1^v$. $x_k$ is the state vector of the voltage model. 
    The inequality \eqref{eq:mpc5} defines the BESS voltage limits.
    The TTC model for computing the DC voltage and the estimation of $x_k$ are described in \cite{sossan_achieving_2016}.
    
    \item The equality \eqref{eq:mpc6} is the evolution of the BESS SOC as linear function of the variable $\bm{i}_{\overline{k}|k}^o$, where $\phi^{\text{SOC}}$ and $\psi_i^{\text{SOC}}$ are transient matrices obtained from the BESS SOC model: 
    \begin{equation}
    SOC_{k+1} = SOC_k +\frac{10}{3600}\frac{i_k}{C_{nom}} \label{eq:SOCssm}
    \end{equation}
where $C_{nom}$ is BESS capacity in [Ah]. The discritized state-space matrix for the SOC model can be easily obtained from~\eqref{eq:SOCssm} with $A_s=1$, $B_s=10/3600/C_{nom}$, $C_s=1$, $D_s=0$. Finally, \eqref{eq:mpc7} enforces the limits on BESS SOC.
\end{itemize}
The optimization problem is solved at each time step $k$ obtaining the control trajectory for the whole residual horizon from the index $k$ to $\overline{k}$, i.e., $\bm{i}_{\overline{k}|k}$. However, only the first component of the current control trajectory is considered for actuation, i.e.,  $\bm{i}_{\overline{k}|k}^o$. Then, $\bm{i}_{\overline{k}|k}$ is transformed into a power set-point $P^{0}_{k}$, computed as:
\begin{equation}
    P^{0}_{k} = v_k\cdot i_k^o \label{eq:MPCsetpoint}
\end{equation}


\subsection{Real-Time Control Stage}\label{section:RTcontroller}
The real-time control stage is the final stage of the framework, whose output $f_{ref}$, $v_{ref}$ is the input for the GFR BESS converter.
Thanks to the day-ahead problem, sufficient BESS energy capacity is guaranteed in the MPC tracking problem. To ensure the BESS operation to be within the power limits, a static physical constraint of control actions is considered in the day-ahead stage in \eqref{eq:constraint3} and \eqref{eq:constraint4} and during the dispatch tracking in \eqref{eq:mpc2} and \eqref{eq:mpc3}. Nevertheless, these constraints do not account for the dependency of the converter feasible PQ region on DC voltage and AC grid voltage conditions since they are only known in reality. In this respect, the real-time controller is implemented to both keep the converter operating in the PQ feasible region identified by the capability curve and to convert the power set-point from the MPC problem into a frequency reference set-point to feed the GFR converter.
\subsubsection{Capability Curve}
As proved in \cite{zecchino_optimal_2021}, the converter PQ capability curve $h$ can be modelled as a function of the BESS DC voltage $v_t^{DC}$ and the module of the direct sequence component of the phase-to-phase BESS AC side voltages $v_t^{AC}$ at time $t\in [1,2,...T]$ as:
\begin{equation}
    h(P_t,Q_t,v_t^{DC},v_t^{AC},SOC_t)\leq 0 \label{eq:pqcurve1}
\end{equation}
where the BESS SOC is considered for the selection of the capability curve because the estimation of $v_t^{DC}$ relies on the battery TTC model whose parameters are SOC-dependent~\cite{zecchino_optimal_2021}. 
In particular, the $v_t^{DC}$ is estimated using the TTC model of DC voltage, thus, the same formula as~\eqref{eq:mpc4}: 
\begin{equation}
 v_{t}^{DC} = \phi_v x_{t} +\psi_i^v{i}_t^{DC}+\psi_1^v\bm{1}  \label{eq:estVdc}
\end{equation}
Equation (\ref{eq:estVdc}) is solved together with the charging or discharging DC current equation as follows:
\begin{equation}
    {i}_t^{DC} \approx \frac{{P}_t^{DC}}{v_{t}^{DC} } \label{eq:estIdc}
\end{equation}
where the active power at the DC bus is related to the active power set-point AC side of the converter as:
\begin{equation}
    P_t^{DC} = \left\{\begin{array}{cc}
         \eta P_{set,t}, & \forall P_{set,t} < 0  \\
         P_{set,t}/\eta , & \forall P_{set,t} \leq 0
    \end{array}  \right.  \label{eq:Pac2Pdc}
\end{equation}
where $\eta$ is the efficiency of converter, $P_{set,t}$ is the set-point from the MPC, computed in~\eqref{eq:MPCsetpoint} and expressed as:
\begin{equation}
    P_{set,t} = P_{\floor{\frac{t}{10}}}^{0}
\end{equation}
Once the DC voltage $v_t^{DC}$ is known, the magnitude of the direct sequence component $v_t^{AC}$ of the phase-to-phase voltage at AC side of the converter is estimated via the Thévenin equivalent circuit of the AC grid, expressed as 
\begin{equation}
    v_t^{AC} \approx\sqrt{(v_t^{AC,m})^2+X_T^2\frac{(P_{set,t})^2+(Q_{set,t})^2}{3(v_t^{AC,m})^2}} \label{eq:EstVac}
\end{equation}
where the primary side voltage $v_t^{AC,MV}$ is referred to the secondary side as  $v_t^{AC,m} = v_t^{AC,MV}\frac{1}{n}$, being $n$ the transformer ratio, $v_t^{AC,MV}$ is the voltage measured at the primary side of the transformer, and $X_T$ is the reactance of the step-up transformer, as shown in Fig.~\ref{fig:ExperimentalSetup}. Equations \eqref{eq:pqcurve1} to \eqref{eq:EstVac} represent the relation between active and reactive power set-points with the converter capability curve.

\subsubsection{Set-point conversion for GFR converters}
Together with a feasibility check for the power set-point $P_{set,t}$, the real time controller is responsible for converting the power set-point into a frequency reference set-point to feed the GFR converter. In particular, the power output of a GFR converter can be expressed, starting from \cref{eq:GFRfref}, as:
\begin{equation}
    P =\sigma_f \cdot (f-f_{\text{nom}})+ \sigma_f \cdot( f_{\text{nom}}-f_{ref,t}) =  P_{fcr} + P_{set,t}
    \label{eq:GFRfref}
\end{equation}
where $f_{nom}$ is the nominal frequency, the term $P_{fcr}$ corresponds to the power delivered with respect to the frequency containment action, $f$ is the grid frequency\footnote{It should be noted that the frequency control action $P_{fcr}$ and the grid frequency $f$ are not denoted with subscript $t$ because they are not controlled variables in the optimization problem. Instead, $f$ depends on the interconnected power grid and $P_{fcr}$ is the automatic response of GFR control with response time in the order of tens of milliseconds.}. 
As visible from Equation (\ref{eq:GFRfref}), the relation between the power set-point $P_{set}$ of the GFR converter and the input $f_{ref}$  is linear:
\begin{equation}
    P_{set,t} = \sigma_f \cdot (f_{\text{nom}}-f_{ref,t})
\label{eq:GFRfref2}
\end{equation}
Similarly, for the reactive power: 
\begin{equation}
    Q_{set,t} = \sigma_v \cdot (v_{\text{nom}}-v_{ref,t})
\label{eq:GFRvref2}
\end{equation}
where 
$v_{\text{nom}}$ is the nominal voltage.
Equations \eqref{eq:GFRfref2} to \eqref{eq:GFRvref2} represent the relation between active and reactive power with frequency and voltage set-point fed to a GFR converter.

\subsubsection{Real-time problem formulation}
Finally, given a set-point in power $P_{set,t},Q_{set,t}$ (in order to prioritize the active power, the reactive power set-point $Q_{set,t}$ can be set as zero) coming from the MPC problem, the GFR converter optimal references are computed by solving the following optimization problem:
\begin{equation}
  \begin{aligned}
& [f_{ref,t}^*, v_{ref,t}^*] =  \\
& = \!\argmin
\lambda_P(P_{set}^* - P_{set,t})^2 + \lambda_Q(Q_{set}^* - Q_{set,t})^2       \end{aligned}
\label{eq:OPT_RT}
\end{equation}
subject to (\ref{eq:pqcurve1}) - (\ref{eq:GFRvref2}), where (\ref{eq:pqcurve1}) - (\ref{eq:EstVac}) represent the relation between active/reactive power with the converter capability curve and  (\ref{eq:GFRfref2}), (\ref{eq:GFRvref2}) represent the relation between active/reactive power with frequency/voltage set-point fed to the GFR controller.
A way to convexify the problem in \eqref{eq:OPT_RT} subject to (\ref{eq:pqcurve1}) - (\ref{eq:EstVac}) has been presented in \cite{zecchino_optimal_2021}, while constraints \eqref{eq:GFRfref2} - \eqref{eq:GFRvref2} are linear, since $\sigma_{f}$ and $\sigma_{v}$ are fixed.
The optimization problem is defined to find the optimal active and reactive power set-point compatibly with the capability curve of the converter. In particular, if the original set-points are feasible, the optimization problem returns the obvious solution  $P_{set}^*=P_{set,t}$ and the converter reference points are:
\begin{align}
    f_{ref}^* &= f_{\text{nom}} - \frac{P_{set}^*}{\sigma_f} \\
    v_{ref}^* &= v_{\text{nom}} - \frac{Q_{set}^*}{\sigma_v}
\end{align}


        
\section{Results}
\label{Section:Results}
\subsection{Experimental Setup}
\label{SubSection:Expsetup}

For the experimental campaign, a 20 kV distribution feeder in the EPFL campus equipped with a BESS is considered, as shown in 
\cref{fig:ExperimentalSetup}. The distribution feeder includes a group of buildings characterised by a  140 kW base load, hosting 105 kWp root-top PV installation and a grid-connected 720 kVA/500 kWh Lithium Titanate BESS. The targeted grid has a radial topology and is characterized by co-axial cables lines with a cross section of 95 mm$^2$ and a length of few hundreds meters, therefore, the grid losses are negligible \cite{pignati_real-time_2015}.
The measuring systems is composed by a Phasor Measurement Unit (PMU)-based distributed sensing infrastructure.  The measuring infrastructure allows for acquiring in real time accurate information of the power flows $G$, $L$ and $P$, thanks to the PMUs' fast reporting rate (i.e., 50 frame per second) and high accuracy which in terms of 1 standard deviation is equal to $0.001$~degrees (i.e. 18~$\mu$rad) \cite{PSCC_lorenzoLausanne}.

\subsection{Experimental Validation}
This subsection reports the results of a day-long experiment, taking place on the EPFL campus on a working day (Friday, i.e., day-category C according to \cref{Append:Forecasting}). 

\subsubsection{Day Ahead}
\begin{figure}[t]
    \centering
    \subfloat[Day-ahead load scenarios]
    {%
        \centering
        \includegraphics[width=0.9\linewidth]{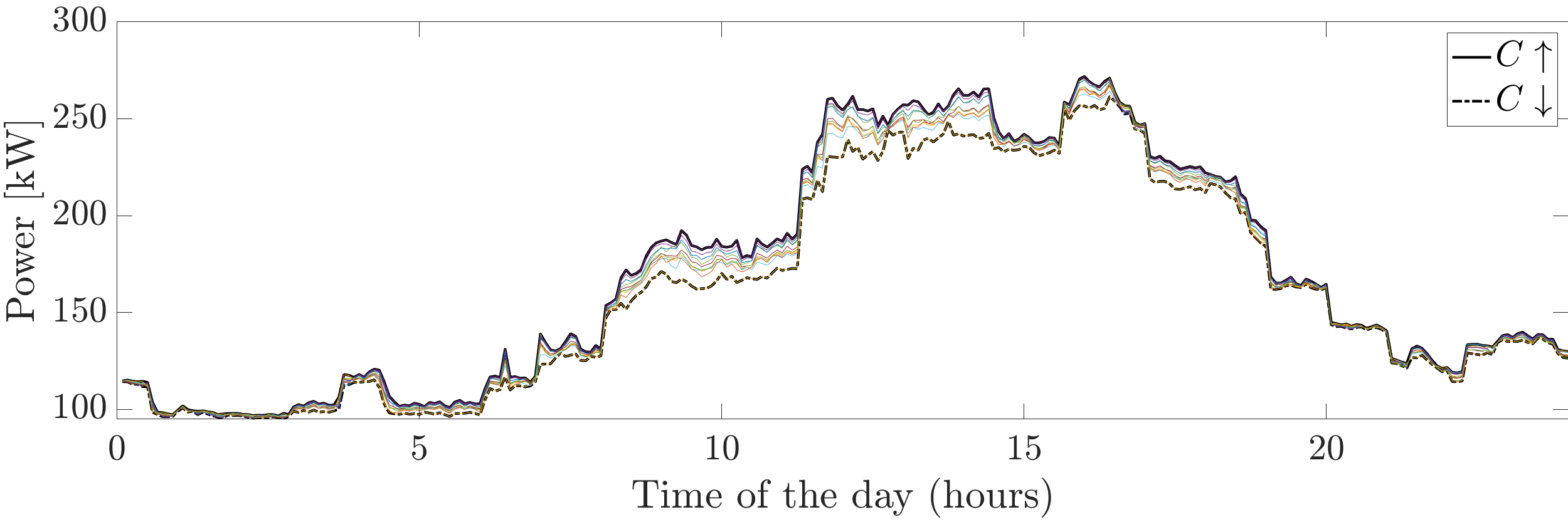}
        \label{fig:dayahead_load}
    } \\
    \subfloat[Day-ahead PV forecast]
    {%
        \centering
        \includegraphics[width=0.9\linewidth]{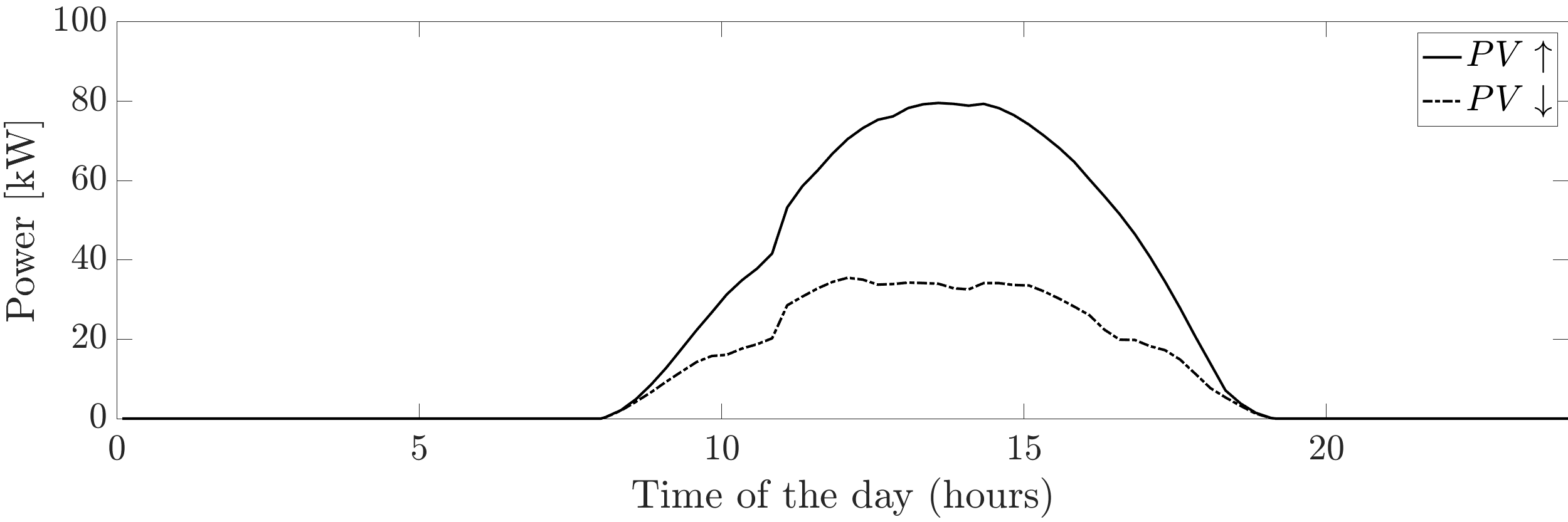}
        \label{fig:dayahead_PV}
    } \\
    \subfloat[Day-ahead net demand scenarios]
    {%
        \centering
        \includegraphics[width=0.9\linewidth]{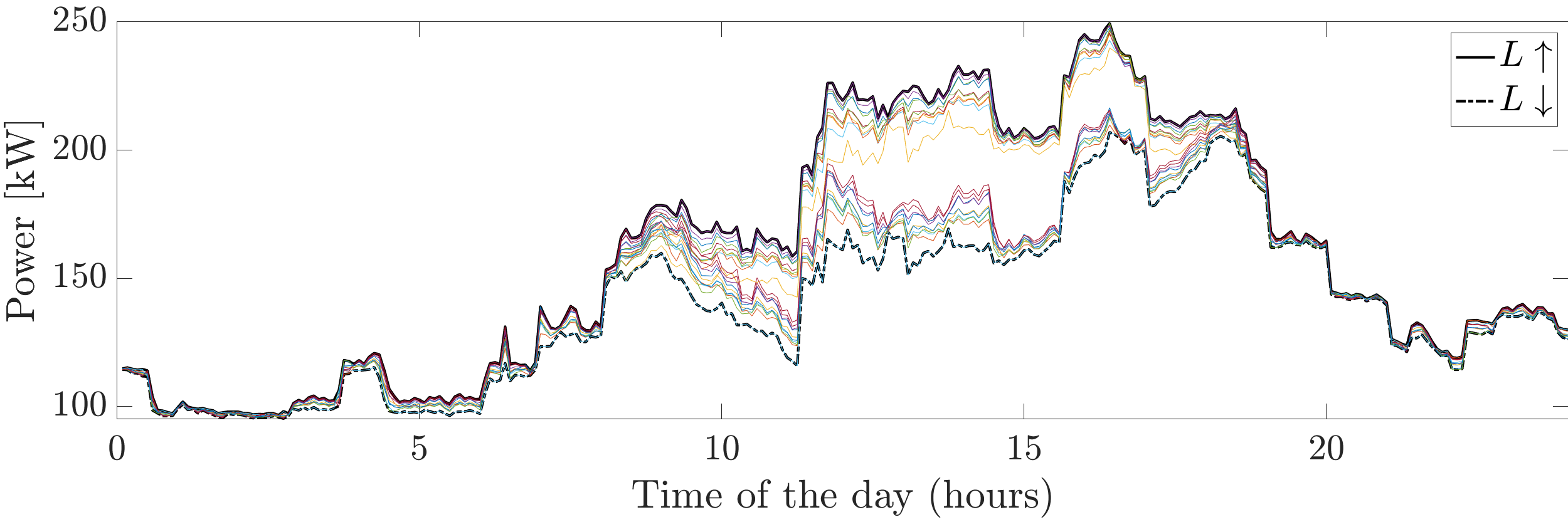}
        \label{fig:dayahead_prosum}
    } \\
    \subfloat[BESS power budget]
    {%
        \centering
        \includegraphics[width=0.9\linewidth]{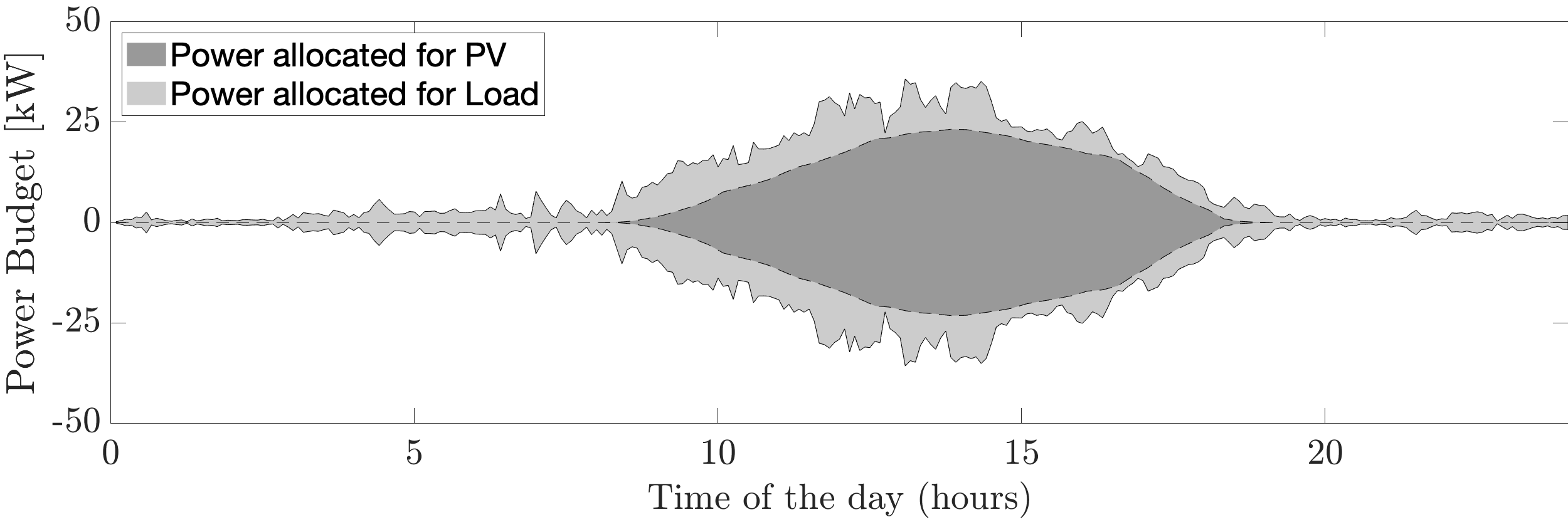}
        \label{fig:dayahead_Pbud}
    } \\
    \subfloat[BESS energy budget]
    {%
        \centering
        \includegraphics[width=0.9\linewidth]{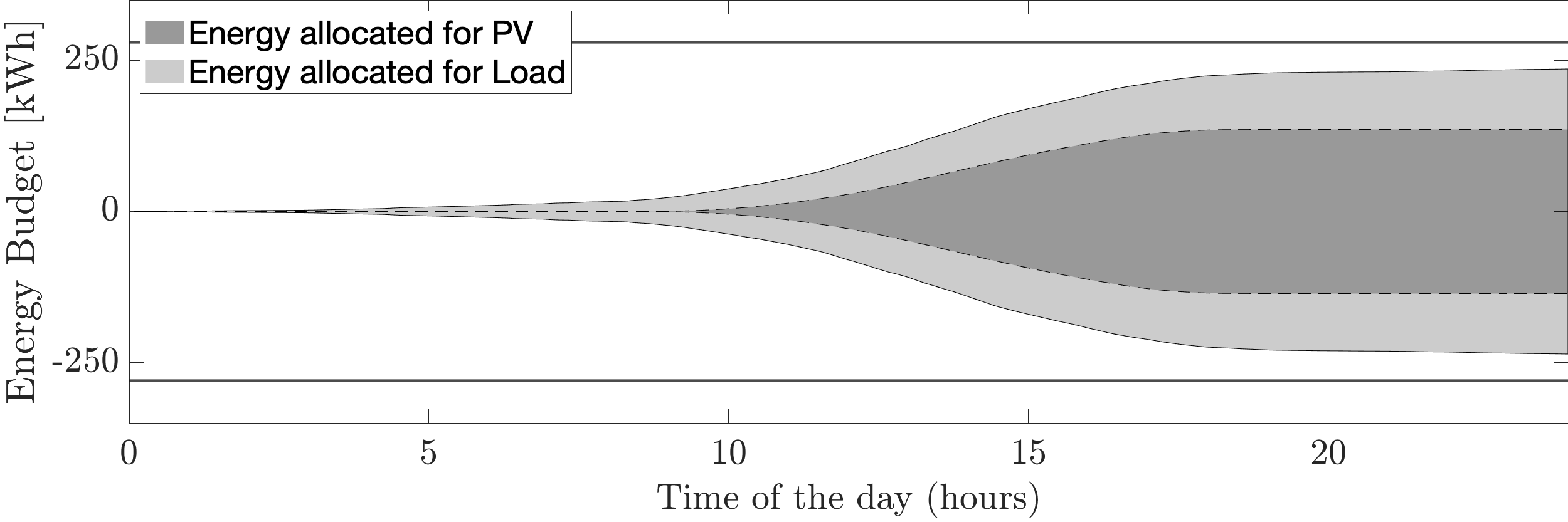}
        \label{fig:dayahead_Ebud}
    } \\
       \caption
    {
        Input and output of the day-ahead problem. (a) shows 10 demand generated scenarios, and their relative upper and lower bound. (b) shows the upper and lower bound for the PV production. (c) combines load and PV to show the prosumption scenario, input of the day ahead problem. (d) and (e) show the power and energy budget allocated to perform the different services. 
    }
    \label{fig:dayahead_results}
\end{figure}

The input and output information of the day-ahead dispatch process for the experimental day are shown in Fig. \ref{fig:dayahead_results}. The $S=10$ generated scenarios $\boldsymbol{C}$ for the prosumption are shown in \cref{fig:dayahead_load}, where $\boldsymbol{C}^{\downarrow}$ and $\boldsymbol{C}^{\uparrow}$ are the lower and upper bounds shown in thick black lines, while all the scenarios are represented by thin colored lines.
The upper and lower bound of the PV forecast, expressed in terms of PV production in kW, are shown in 
\cref{fig:dayahead_PV}, while the net demand scenarios at the PCC, obtained according to \cref{eq:pros_up,eq:pros_down}, are shown in \cref{fig:dayahead_prosum}. The upper and lower bound of the prosumption, namely $   L_n^{\uparrow}$ and $L_n^{\downarrow}$, are inputs to the dispatch plan. Finally, Fig. (\ref{fig:dayahead_Pbud}) and (\ref{fig:dayahead_Ebud}) show respectively the power and energy budget allocated for the forecasting uncertainty of the stochastic PV production (in dark gray) and demand (in light gray). The remaining energy budget is allocated for the FCR  service, resulting in a droop $\sigma_f=116$ kW/Hz.

\subsubsection{Dispatch tracking}
\begin{figure}[t]
    \centering
    \subfloat[Day-ahead load scenarios]
    {%
        \centering
        \includegraphics[width=0.9\linewidth]{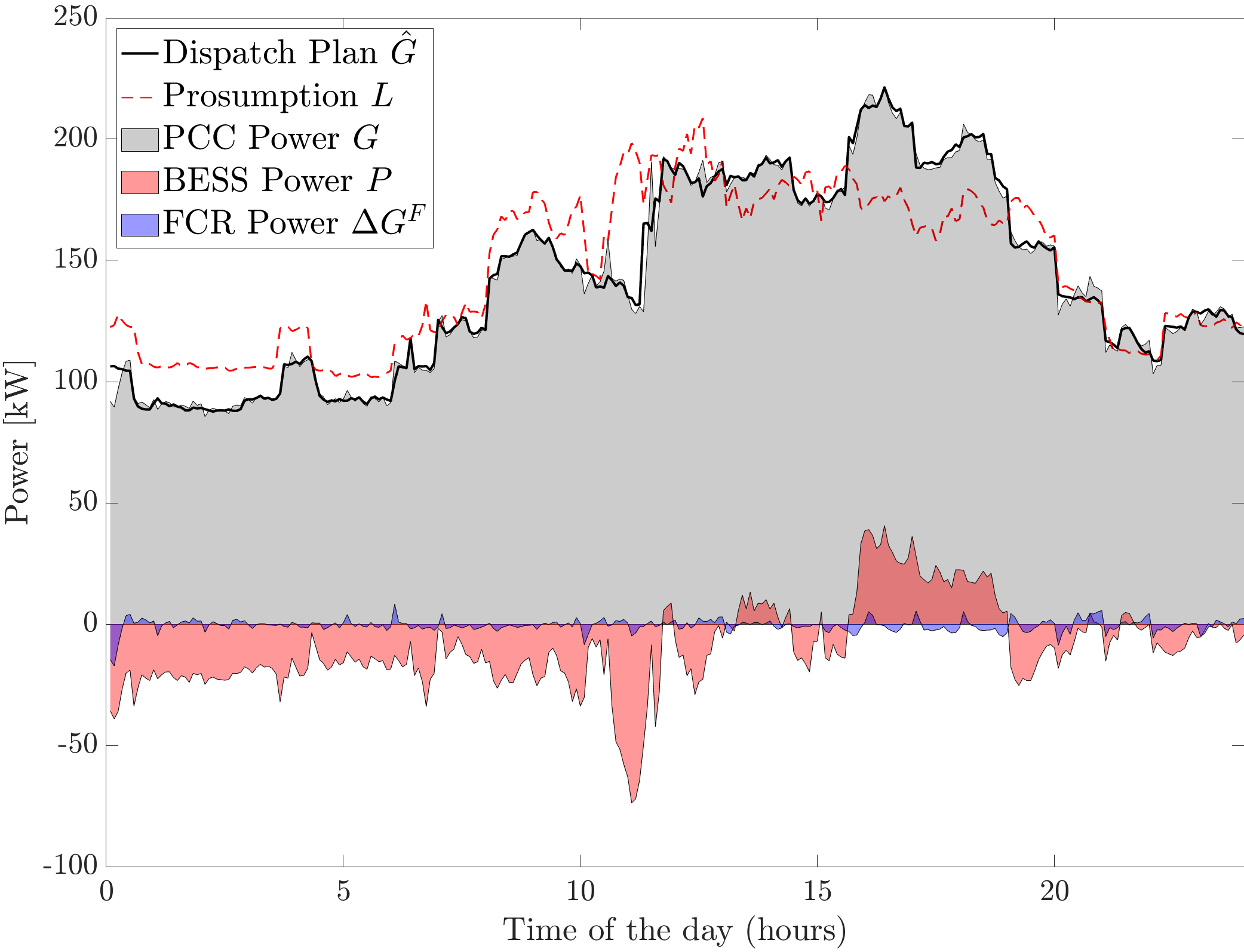}
        \label{fig:experiment}
    } \\
    \subfloat[Day-ahead PV forecast]
    {%
        \centering
        \includegraphics[width=0.9\linewidth]{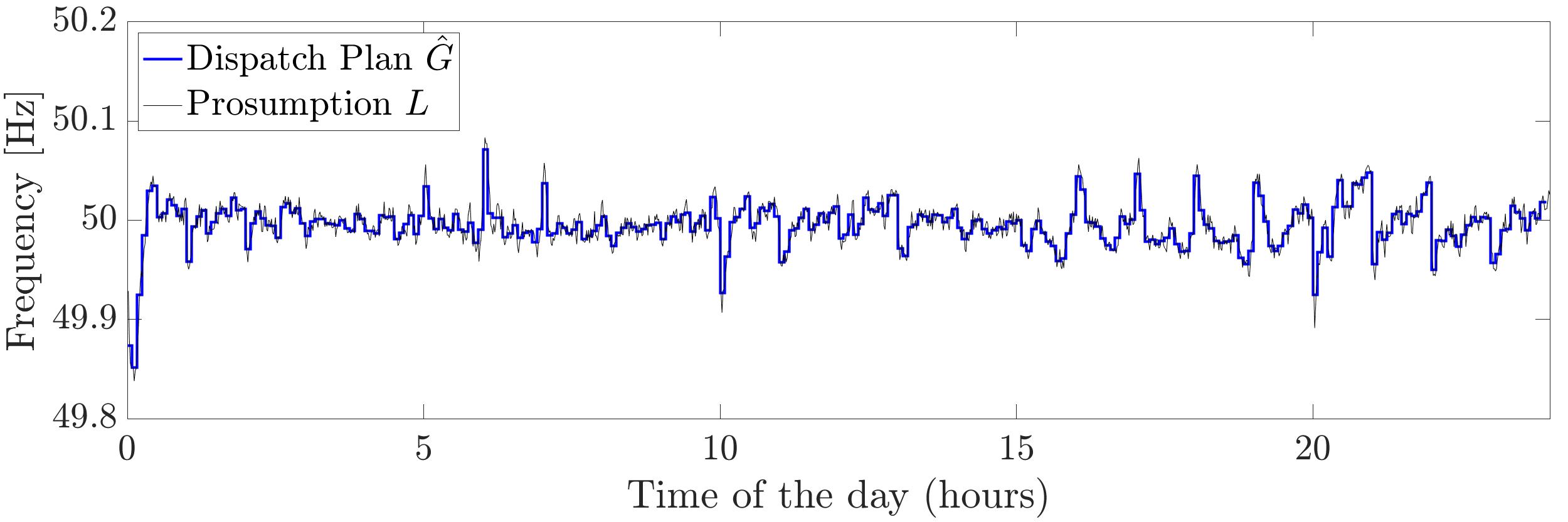}
        \label{fig:experiment_f}
    } \\
    \subfloat[Day-ahead net demand scenarios]
    {%
        \centering
        \includegraphics[width=0.97\linewidth]{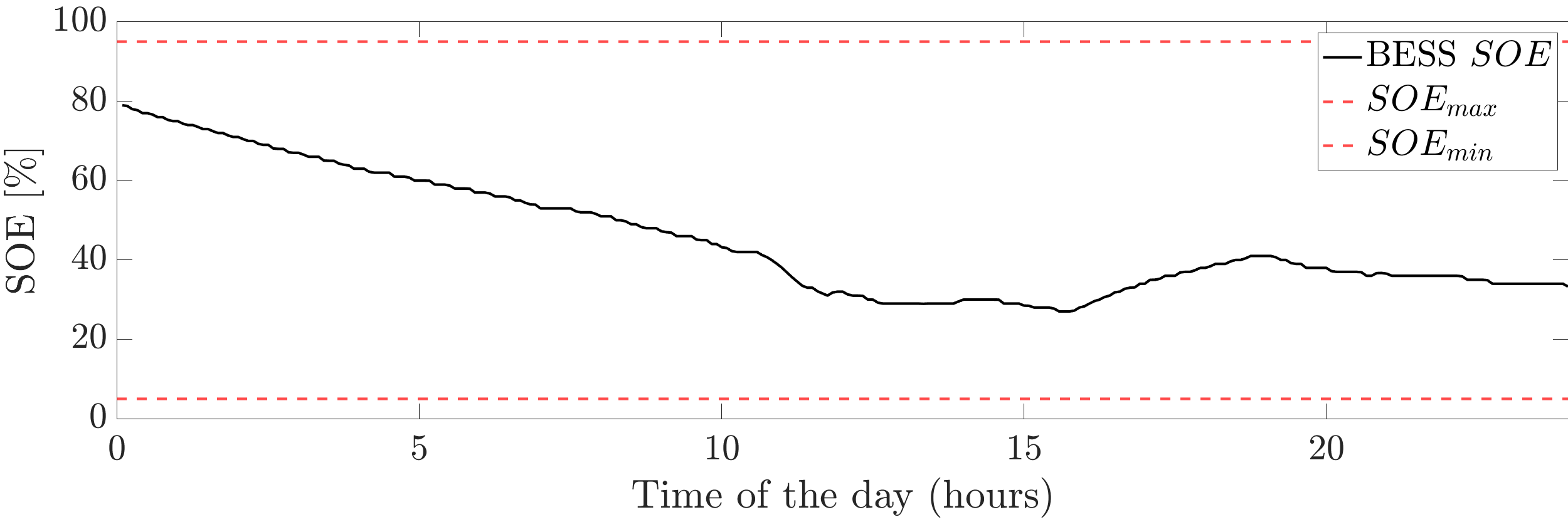}
        \label{fig:experiment_SOE}
    } \\
       \caption
    {
    Experimental results for the 24-hour test. (a) compares the dispatch plan (in black) with: the measured power at the PCC (shaded gray), the prosumption of the feeder (in dashed red), the BESS power flow (in shaded red) and the average power required for each 5-minutes window for the provision of FCR service by the BESS. (b) shows the grid-frequency and its 5-minutes mean. The SOE of the battery during the test is visible in (c).
    }
    \label{fig:experiment_results}
\end{figure}

The results of the dispatch tracking are visible in Fig. \ref{fig:experiment_results}. In particular, Fig. \ref{fig:experiment} shows the power at the PCC (in shaded gray), the prosumption (in dashed red) and the dispatch plan (in black). It is observed that the dispatch plan is tracked by the GFR-converter-interfaced BESS when the grid-frequency (visible in \cref{fig:experiment_f}) is close to 50 Hz. Nonetheless, when the grid-frequency has a significant deviation from 50 Hz, the GFR converter provides a non negligible amount of power $\Delta G^F$ to the feeder. This contribution is visible in shaded blue in \cref{fig:experiment}, and causes a deviation of the average PCC power from the dispatch plan, as targeted by \cref{eq:MPCobj}. Moreover, as visible in \cref{fig:experiment_SOE} the BESS SOE is contained within its physical limits over the day. 
\begin{table}[]
    \renewcommand{\arraystretch}{1.3}
    \centering
    \caption{TRACKING ERROR STATISTICS (IN KW)}
    \resizebox{0.7\linewidth}{!}{%
\begin{tabular}{@{}lccc@{}}
\toprule
                        & ME  & MAE     & RMSE  \\ \midrule
No Dispatch Tracking    & -3.49 & 47.00   & 18.26 \\
Dispatch Tracking       & 0.11  & 16.93   & 3.00  \\
Dispatch + FCR Tracking & -0.45 & 0.79    & 1.43  \\ \bottomrule
\end{tabular}}
\label{table:I}
\vspace{-0.5cm}
\end{table}
To evaluate the dispatch plan-tracking performance, \emph{Root Mean Square Error} (RMSE), \emph{Mean Error} (ME), and \emph{Maximum Absolute Error} (MAE) are considered. In particular these indicators are visible in \cref{table:I} for three different cases:
\begin{enumerate*}[label=(\roman)]
    \item[(i)] no dispatch case, where the error is computed as difference between prosumption and dispatch plan;
    \item[(ii)] dispatch tracking case, where the error is computed as difference between flow at the PCC and dispatch plan;
    \item[(iii)] dispatch tracking + FCR case, where the error is computed as difference between flow at the PCC and dispatch plan + FCR contribution, as targeted by the MPC problem (\ref{eq:mpcProb}).
\end{enumerate*}
The obtained results are proving the good performance of the dispatch + FCR tracking framework. The overall results are comparable with the one presented in \cite{sossan_achieving_2016}. 
Nevertheless, it must be noticed that in this study, the tracking problem is complicated by the fact that the BESS is operated with a grid-forming converter providing frequency regulation.

\subsubsection{Frequency Regulation}
To assess the performance of the GFR converter in regulating the frequency at the PCC, we adopt the metric relative Rate-of-Change-of-Frequency (rRoCoF) proposed in \cite{zecchino_local_2021}, defined as:
\begin{align}
rRoCoF = \left | \frac{\Delta f_{PCC} / \Delta t}{\Delta P}\right |
\end{align}
where  $\Delta f_{PCC}$ is the difference between one grid-frequency sample and the next (once-differentiated value) at the PCC, $\Delta P$ is the once-differentiated BESS active power, and $\Delta t$ is the sampling interval. As the metric rRoCoF is weighted by the delivered active power of the BESS, it can also be used to compare the effectiveness of converter controls (i.e.,GFR vs GFL) in regulating frequency at local level in a large inter-connected power system. 
The grid-frequency is measured by a PMU installed at the PCC.

The rRoCoF is computed from different frequency timeseries, corresponding to the following four cases.
\begin{itemize}
    \item \textit{Case 1}: rRoCoF computed considering the 24 hour-long experiment where GFR-controlled BESS is providing multiple services. 
    
    \item \textit{Case 2}: rRoCoF computed considering a 15-minute window around a significant frequency transient (i.e.,around 00:00 CET) during the same day-long experiment.
    
    \item \textit{Case 3}: rRoCoF computed with a dedicated 15-minute experiment where GFR-controlled BESS is only providing FCR with its highest possible frequency-droop (1440 kW/Hz) during a significant grid-frequency transient.
    
    \item \textit{Case 4}: rRoCoF computed with a dedicated 15-minute experiment where GFL-controlled BESS is only providing FCR with its highest possible frequency-droop (1440 kW/Hz) during a significant grid-frequency transient.
\end{itemize}
While \textit{Case 1} and \textit{Case 2} rely on the measurements obtained from the experiment carried out in this study, \textit{Case 3} and \textit{Case 4} leverage an historical frequency data-set, also used in the experimental validation proposed in \cite{zecchino_local_2021}. It should be noted that the same experimental setup described in~\cref{SubSection:Expsetup} is utilized in \cite{zecchino_local_2021}. The measurements at hour transition are considered in order to evaluate GFR/GFL units' frequency regulation performance under relatively large frequency variations.  
\begin{figure}[]
    \centering
    \includegraphics[width=1\linewidth]{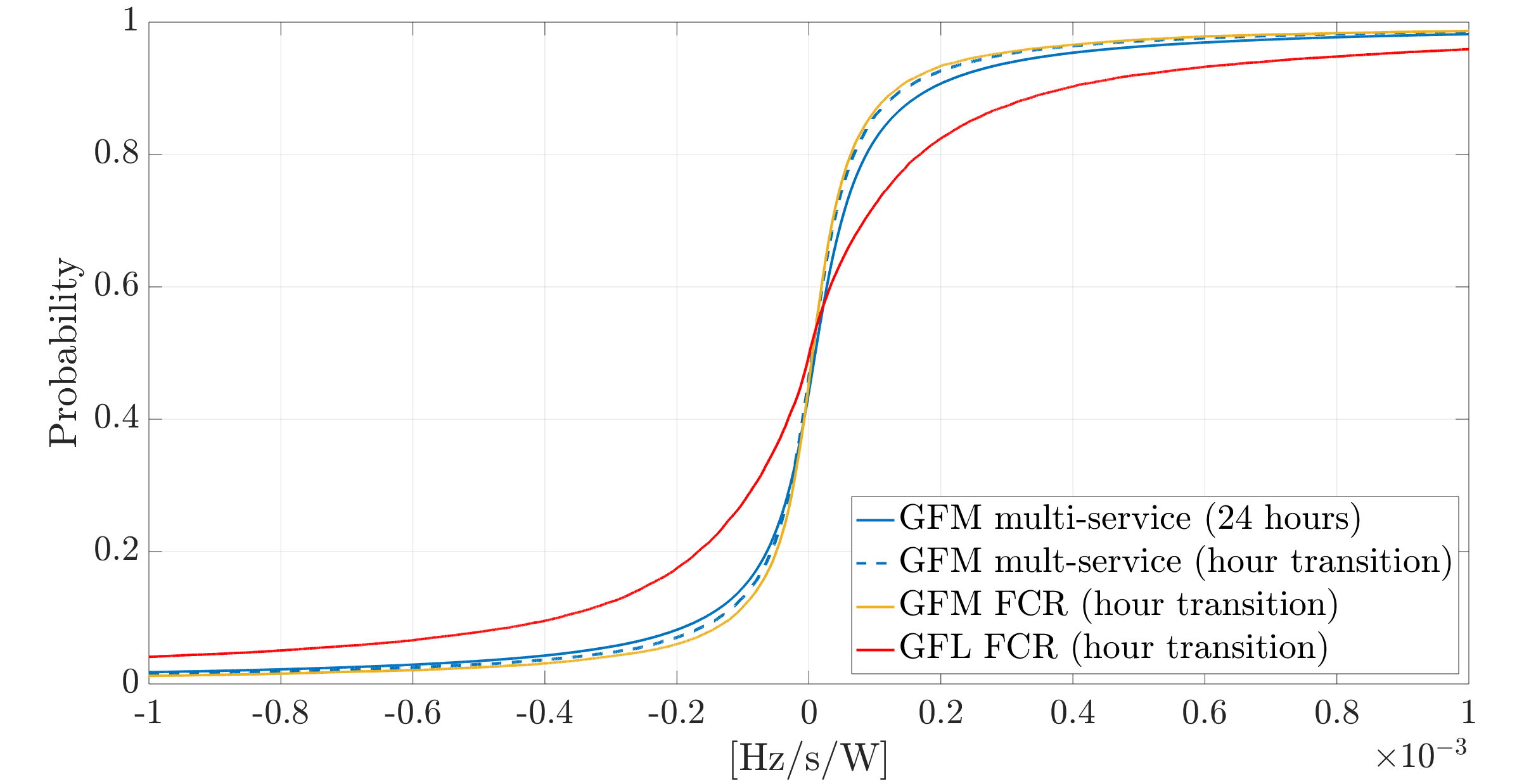}
    \caption{Cumulative Density Function of rRoCoF.}
    \label{fig:rRoCoF}
\end{figure}

\cref{fig:rRoCoF} shows the \emph{Cumulative Density Function} (CDF) of the rRoCoF values for the four cases. 
First, it is observed that the CDF results of \textit{Case 1} and \textit{Case 2} are very close. This illustrates the robust of the metric rRoCoF. In particular since rRoCoF is normalised by the BESS power injection, the frequency dynamic and the frequency droop of the BESS controller have little effect on the result of CDF. Moreover, the comparison between \textit{Case 2} and \textit{Case 3} shows negligible differences on the rRoCoF CDF, proving that the provision of dispatchability service by the GFR converter does not drastically affect its frequency regulation performance. 
Finally, the comparison between GFR and GFL-controlled BESS (i.e.,  \textit{Case 2} vs \textit{Case 4}) is reported from \cite{zecchino_local_2021}, to show that GFR unit achieves significantly lower rRoCoF for per watt of regulating power injected by the BESS. 
        
\section{Conclusions}
\label{Section:Conclusions}
A comprehensive framework for the simultaneous provision of multiple services (i.e. feeder dispatchability, frequency and voltage regulation) to the grid by means of a GFR-converter-interfaced BESS has been proposed.
The framework consists of three stages. The day-ahead stage determines an optimal dispatch plan and a maximum frequency droop coefficient by solving a robust optimization problem that accounts for the uncertainty of forecasted prosumption. In the intra-day stage, a MPC method is used in the operation process to achieve the tracking of dispatch plan while allowing for frequency containment reserve properly delivered by the BESS. Finally, the real-time controller is implemented to convert the power set-points from MPC into frequency references accounting for the PQ feasible region of the converter.

The experimental campaign carried out in a 20 kV distribution feeder in the EPFL campus. The feeder includes a group of buildings characterised by a 140 kW base load, hosting 105 kWp root-top PV installation and a grid-connected 720 kVA/500 kWh Lithium Titanate BESS. 
A 24-hour long experiment proved good performance in terms of dispatch tracking, compatible with the ones obtained in \cite{sossan_achieving_2016} for sole provision of dispatchability by means of GFL converter-interfaced BESS.
Moreover, the rRoCoF metric has been used to shows that the provision of dispatchability service by the GFR converter does not affect its frequency regulation performance and confirm the positive effects of grid-forming converters with respect to the grid-following ones in the control of the local frequency.
Future works concern the development of control strategies to prioritize the FCR service when the battery is operating close to
its operational limit, as required by grid-codes. Moreover, an analysis on the effects of voltage regulation on the BESS losses could be included in the day ahead scheduling problem.
\section*{Acknowledgement}
This work is supported by OSMOSE project. The project has received funding from the European Union's Horizon 2020 Research and Innovation Program under Grant 773406.
\begin{appendices}
\section{Forecasting Tools}
\label{Append:Forecasting}
Once the optimization problem \cref{eq:optProb,eq:constraint1,eq:constraint2,eq:constraint3,eq:constraint4} is defined, the challenge stands in the forecast of the prosumption and the frequency deviation, in particular their confidence intervals, denoting the maximum and minimum expected realisations, namely $L_n^\uparrow$ , $L_n^\downarrow $ and $W_{f,k}^\uparrow$, $ W_{f,k}^\downarrow $.
As correctly appointed by \cite{sossan_achieving_2016}, the local prosumption is characterized by a high volatility due to the reduced spatial smoothing effect of PV generation and the prominence of isolated stochastic events, such as induction motors inrushes due to the insertion of pumps or elevators. 
For these reasons, the existing forecasting methodologies, developed by considering high levels of aggregation, e.g. \cite{chen_load_2004}, are not suitable to predict low populated aggregates of prosumers.
For the proposed application, the problem of identifying $L_n^\uparrow$ and $L_n^\downarrow $ is divided into two sub-problems:
\begin{enumerate*}[label=(\roman*)]
    \item load consumption forecast and
    \item PV production forecast.
\end{enumerate*}
For the first one, a simple non-parametric forecasting strategy relying on the statistical properties of the time series is proposed. The PV production forecast is performed by taking advantage of solar radiation and meteorological data services providing a day-ahead prediction of the \emph{Global Horizontal Irradiance} (GHI) together with its uncertainty. The GHI forecast, together with the information related to the PV installation (i.e. total capacity, location, tilt, azimuth) allows computing the \emph{Global Normal Irradiance} (GNI) and obtaining an estimation of the total PV production, and the related uncertainties. The best and worst PV production scenarios are computed by transposing the GHI forecast data and applying a physical model of PV generation accounting for the air temperature \cite{sossan_solar_2019}. For a given day-ahead forecast, the vector containing the best and worst production scenario for the PV, with a time resolution of 5 minute are named as $\bf{PV}^{\uparrow}$ and $\bf{PV}^{\downarrow}$, respectively.

As previously mentioned, while the PV production forecast leverages GHI data, the load forecast only relies on statistical properties of recorded time-series. In particular, a set of historical observation $\bm{\mathcal{G}}$ at the PCC point is considered. The historical load consumption $\bm{\mathcal{C}}$ is computed for every time step $n$ corresponding to a 5 minutes window and every day $d$, as:
\begin{equation}
\begin{aligned}
  &  \mathcal{C}_{n,d} = \mathcal{G}_{n,d} - \mathcal{P}_{n,d} - \mathcal{PV}_{n,d} \\ & \forall n \in [1, N] \quad \quad  \forall d  \in [1,D] &
    \label{eq:disaggregation}
\end{aligned}
\end{equation}
where $\mathcal{P}_n$ is the historical measure of the BESS power at time $n$ and $\mathcal{PV}_n$ is the estimated PV production at the same time relying on the onsite measures of GHI, and $D$ is the total number of recorded days.
The process described by \cref{eq:disaggregation}, also know as disaggregation, allows for the decoupling of the PV production and the load consumption $\mathcal{C}$, composed by 288 samples for each recorded day.
The different consumption scenarios are generated by applying the following heuristic:
\begin{itemize}
\item The data-set $\bm{\mathcal{C}}$ is divided into sub-sets $\Omega_{A,B,C,D1,D2}$ by selecting consumption data corresponding respectively to:
    \begin{enumerate*}[label=(\Alph*)]
    \item first working day of the week, i.e. Mondays or days after holidays;
    \item central working days of the week, i.e. Tuesdays, Wednesdays and Thursdays;
    \item last working day of the week, i.e. Friday or days before holidays;
    \item holidays, i.e. Saturday (subcategory D1), Sundays and festivities (subcategory D2).
    \end{enumerate*}
\item For each sub-set, the statistical properties $\mu$ and $\sigma$ are inferred as:
\begin{align}
      \boldsymbol{\mu}_{\Omega} &= \text{\texttt{mean}}(\boldsymbol{\Omega})  \quad \quad
      \boldsymbol{\sigma}_{\Omega} = \text{\texttt{cov}}(\boldsymbol{\Omega})  \\
    \forall \boldsymbol{\Omega} &\in [\Omega_{A,B,C,D1,D2}]  \nonumber
\end{align}
where the function \texttt{mean} returns an array of 288 points, each of which represents a mean value for a particular 5-minutes window of the day and the function \texttt{cov} returns a 288x288 matrix corresponding to the covariance matrix of the observation.
\item Both $\boldsymbol{\mu}_{\Omega}$ and $\boldsymbol{\sigma}_{\Omega}$ are computed by considering an exponential forgetting factor to prioritise the latest measurement with respect to the older one, as defined in \cite{pozzi_exponential_2012}.
\item A given number $S$ of possible scenarios is generated by considering the same multivariate normal distribution, with mean equal to $\mu$ and covariance equal to $\sigma$
\item $\boldsymbol{C}^{\downarrow}$ and $\boldsymbol{C}^{\uparrow}$ are defined as the load scenarios characterised by the lowest and highest load profile, respectively.
\end{itemize}
Finally, the prosumption minimum end maximum expected realisation are computed by combining $\boldsymbol{C}^{\downarrow}$ and $\boldsymbol{C}^{\uparrow}$ with $\boldsymbol{PV}^{\uparrow}$ and $\boldsymbol{PV^{\downarrow}}$ as follows:
\begin{align}
    L_n^{\uparrow} &= C_n^{\uparrow} - PV_n^{\downarrow}   \label{eq:pros_up}\\
    L_n^{\downarrow} &= C_n^{\downarrow} - PV_n^{\uparrow}  \label{eq:pros_down}\\
    \forall n &\in [1, N] \nonumber
\end{align}
Concerning the prediction of $W_f$, while \cite{namor_control_2019} only relies on the statistical properties of the time series, this paper uses an auto regressive model, as supported by \cite{piero_schiapparelli_quantification_2018} which indicates the possibility to predict $W_f$ to reduce the variance of the forecast in respect to the historical variance  of the time series.

\end{appendices}

\bibliographystyle{IEEEtran}
\bibliography{IEEEabrv,reference}
\end{document}